\documentclass[12pt,article,nofootinbib]{revtex4}
\usepackage{pdfpages}
\usepackage{epsf}
\usepackage{subfigure}
\usepackage{amsmath}
\usepackage{eqnarray}
\usepackage{epstopdf}
\usepackage{mathrsfs}
\usepackage{natbib}
\usepackage{amssymb}
\usepackage{dcolumn}
\usepackage{graphicx}
\usepackage{color}
\usepackage{hyperref}
\usepackage{enumerate}
\usepackage{float}

\usepackage{multirow}
\usepackage{blindtext}

\makeatletter
\newcommand*{\rom}[1]{\expandafter\@slowromancap\romannumeral #1@}
\makeatother
\def\be{\begin{equation}}
\def\ee{\end{equation}}
\def\ba{\begin{eqnarray}}
\def\ea{\end{eqnarray}}

\graphicspath{{images/}}

\begin{document}
\title{Cosmological LTB Black Hole in a Quintom Universe}
\author{Sareh Eslamzadeh$^{1}$}
\email{s.eslamzadeh@stu.umz.ac.ir}
\author{Kourosh Nozari$^{1}$}
\email{knozari@umz.ac.ir\,\,(Corresponding Author)}
\author{J. T. Firouzjaee$^{2}$}
\email{firouzjaee@kntu.ac.ir}

\affiliation{$^{1}$Department of Theoretical Physics, Faculty of Science,
University of Mazandaran, P. O. Box 47416-95447, Babolsar, Iran\\
$^{2}$Department of Physics, K. N. Toosi University of Technology,\\
P. O. Box 15875-4416, Tehran, Iran}

\begin{abstract}
		 		
We study cosmological Lemaitre-Tolman-Bondi (LTB) black hole thermodynamics immersed in a quintom universe.
We investigate some thermodynamic aspects of such a black hole in detail.
We apply two methods of treating particles' tunneling from the apparent
horizons and calculate the black hole's temperature in each method; the results of which are the same.
In addition, by considering specific time slices in the cosmic history, we study the thermodynamic features
of this black hole in these specific cosmic epochs. Also, we discuss the information loss problem and the
remnant content of the cosmological black hole in different cosmic epochs in this context.
We show that approximately in all the cosmic history, the temperature of the black hole's
apparent horizon is more than the temperature of the cosmological apparent horizon.\\

	\end{abstract}

\maketitle
\newpage
\tableofcontents
			
	\section{Introduction}
		
Black holes are living in the expanding universe. To be precise, in our expanding universe there are no asymptotically flat black holes. So, it is necessary to treat the physics and thermodynamics of black holes in an expanding cosmological background. Accordingly, black holes asymptotic to the expanding universe, under the title of ``Cosmological Black Holes", have been the subject of many researches these years. Such black holes leave a series of questions, like: What effects does the cosmic expansion have on the local physics of black holes in the entire cosmic epochs? What effects does the content of the universe leave on the black hole? How should be redefined the physics of black holes based on expanding universe? How should be changed the definitions such as black hole horizon, its singularities, and its mass and thermodynamics in an expanding universe? One of the prior research describing black holes in the Friedmann-Robertson-Walker (FRW) universe is the McVittie's solution \cite{33McV}. After that, solutions like Einstein and Strauss \cite{45Ein}, Vaidya \cite{77Vai}, and Lemaitre-Tolman-Bondi (LTB) \cite{34Tol,47Bon,97Lem} have been introduced. The noticeable point in such a research is the redefinition of the horizons based on local concepts, not based on asymptotically flat conditions; which was suggested by Hayward as trapping horizon \cite{94Hay}, and by Ashtekar and Krishnan as dynamical horizon \cite{02Ash}. Besides the dynamic nature of the LTB metric, the FLRW metric can be modeled as a background and is a special case of the LTB metric. Building upon the properties of the LTB metric, a cosmological black hole can be constructed \citep{10-firouzjaee}, where its singularity and horizon are formed during the collapse \citep{11-firouzjaee}. In Refs. \cite{12Van, 15Far}, one can find helpful reviews on the various horizons like event, Killing, apparent, trapping, isolated, and dynamical horizons.\\
		
After the discovery of positively accelerating expansion of the universe \cite{98Rie, 99Per}, the Dark Energy was introduced as a mysterious component responsible for this positively accelerated expansion. The first suggested candidate for this weird component was the cosmological constant \cite{03Pad} . But, problems of the cosmological constant \cite{89Wei} such as fine-tuning, coincidence, and the essence of being constant caused particle physics to give some new alternatives. Therefore, fields like Quintessence \cite{13Tsu}, K-essence \cite{01Arm}, Tachyon \cite{02Sen}, Phantom \cite{02Cal}, and Quintom \cite{05Fen} were some of the most important subsequent suggestions. If we pay attention to the equation of state parameter, $w_{_{field}}=\frac{p}{\rho}$, as an important quantity for a cosmological component, the Quintom field has a fascinating aspect: it is actually a combination of two fields including a Quintessence field with $w>-1$ plus a Phantom field with $w<-1$. Since the observational data are in the favor of a transition from the quintessence phase to a phantom phase at late time, a mechanism for crossing of the cosmological constant equation of state parameter, that is, $w=-1$, is required. In Ref. \cite{10Set}, one can find some observational and theoretical evidences for the necessity of the Quintom field existence as a suitable candidate for the Dark Energy.\\
	
		The connection between thermodynamics variables and black hole geometry was firstly introduced by Bekenestein \cite{73Bek}. Afterward, four laws of thermodynamics  for black holes were established \cite{73Bar} and, then, Hawking initiated the research on the possibility of black hole evaporation \cite{75Haw}. There are two straightforward approaches to calculate the particle tunneling rate from the black hole horizon: One based on the Hamilton-Jacobi method \cite{99Sri}, and the other based on the null geodesics method \cite{00Par, 04Par}. In Ref. \cite{12Van} and references therein, one can find an elegant review on the topic of tunneling methods and Hawking's radiation from both stationary and dynamical black holes. Besides, thermodynamic features of cosmological black holes have been of interest in some research works~\cite{88Sus, 07Far, 07Kie, 10Sus, 11Cha, 12Fir, 14Far, 22Esla}.  \\
		
		The present study aims to probe the tunneling process from the horizons of the cosmological LTB black hole surrounded by a quintom field. In this regard, in section II, we illustrate spacetime which contains the cosmological LTB black hole in the Quintom field as the background dark energy. We characterize the initial conditions which are required to construct both cosmological and black hole apparent horizons.  Also, we debate on what effects the existence of Quintom has on these horizons in the entire cosmic history. In section III, we apply the Parikh-Wilczek method to calculate the entropy and temperature of the cosmological and black hole apparent horizons. Besides, we investigate the correlation between radiative modes and black hole remnant. In section IV, we are curious about the time evolution of the cosmological black hole surrounded by Quintom matter; precisely their horizons and thermodynamics time evolution in the entire cosmic history. Finally, we summarize our results in section V.

	\section{Cosmological LTB Black Hole in a Quintom Universe}
	
		To construct the metric of the cosmological LTB black hole in the Quintom dominated universe, we benefit the reults of Ref. \cite{11Gao}. In this regard, we assume the line element to be as follows
	\begin{equation}\label{LineElement}
	ds^2=-dt^2+e^{\bar{\phi}} dr^2+e^\phi d\Omega^2 ,
	\end{equation}
where $t$ is a cosmic time parameter and $(r,\theta,\varphi)$ are comoving coordinates with $d\Omega^2=d\theta^{2}+\sin^{2}\theta d\varphi^{2}$; $\phi$ and $\bar{\phi}$ are functions of $t$ and $r$. We consider the energy-momentum tensor of the Quintom field in the perfect fluid form as
	\begin{equation}\label{E-MT}
	T_{\mu\nu}=(\rho+p) u_{\mu} u_{\nu} + p g_{\mu\nu} ,
	\end{equation}
where $\rho$ and $p$ are density and pressure of the Quintom field, respectively; and $u^{\mu}=(1,0,0,0)$ is the four-velocity. Assuming there is no accretion, $G^0_1=0$ (see~\cite{11Gao}), other components of the Einstein's field equations are as follows
	\begin{eqnarray}\label{EinCom}
	G_0^0 &=8\pi \rho,\\
	G_1^1 =G_2^2 =G_3^3 &=-8\pi p. \nonumber
	\end{eqnarray}

	As explained in Ref. \cite{11Gao}, taking into account the source to be a single perfect fluid and the background to be spatially flat, the comoving observer realizes a spatially homogenous pressure. Therefore, the Einstein equations give
	\begin{eqnarray}\label{phieq}
	{\ddot{\phi}}+\frac{3}{4} {\dot{\phi}}^2 &= -8\pi p(t),\\
	\frac{{\dot{\phi}^\prime}{\dot{\phi}}}{{\phi}^\prime}+\frac{3}{4} {\dot{\phi}}^2 &= 8\pi \rho (r,t) , \nonumber
	\end{eqnarray}
where overdot and prime denote differentiation with respect to $t$ and $r$, respectively. Following  Ref. \cite{11Gao}, we set the pressure in the form
	\begin{equation}\label{E-MT}
	p=- \frac{p_0}{(t_0-t)^2} ,
	\end{equation}
	where $p_0$ is a positive constant and $t_0$ is recognized as the Big Rip singularity time. The solution of the Eqs. (\ref{phieq}) is given by

	\begin{eqnarray}\label{Sol1}
	e^{\phi}=\big[P(r) (t_0-t)^{\frac{1-k}{2}}+S(r)(t_0-t)^{\frac{1+k}{2}}\big]^{\frac{4}{3}} ,
	\end{eqnarray}
where $k\equiv\sqrt{1+24\pi p_0}$ is a constant in terms of $p_0$; $P$ and $S$ are arbitrary functions of $r$. By choosing $P=r^{3/2}$, $S$ is determined in such a way that the boundary conditions would be recovered correctly. Finally, the metric functions of the cosmological black hole in a Quintom dominated universe are found as follows \cite{11Gao}

	\begin{equation}\label{Sol02}
	e^{\phi}=\bigg[r^{\frac{3}{2}} (t_0-t)^{\frac{1-k}{2}}-\bigg(\frac{3}{2}\sqrt{2M}+\sqrt{6\pi\rho_0}  r^{\frac{3}{2}}\bigg)(t_0-t)^{\frac{1+k}{2}}\bigg]^{\frac{4}{3}} ,
	\end{equation}
and
	\begin{equation}\label{Sol2}
	e^{\bar{\phi}}=\frac{\phi^{\prime2}}{4} e^{\phi}.
	\end{equation}
	
	To compare and check the boundary conditions, one can find in Ref. \cite{00Cel} the cosmological LTB black hole described with the line element as follows
	
	\begin{equation}\label{LTB}
	ds^2=-dt^2+\frac{R'^2 (r,t)}{1+2 E(r)} dr^2+R^2 (r,t) (d\theta^2+\sin^2 \theta d\varphi^2) ,
	\end{equation}
where $R(r,t)$ is a physical radius, $E(R)=\frac{1}{2} \dot{R}^2(r,t) -\frac{M(R)}{R(r,t)}$ gives the meaning of the total energy per unit mass, while $M(R)$ is the mass in the sphere of comoving radius $r$. If a collapsing metric is built by this metric, one can show that the apparent horizon (trapping horizon or dynamical horizon) will form at $ R = 2M $	surface. The quantity $E(r)$ is like the curvature function which includes a contribution from the kinetic energy and the gravitational potential energy. To investigate the boundary conditions of the metric Eqs. (\ref{Sol02}) and (\ref{Sol2}), we compare Eqs. (\ref{LineElement}), (\ref{Sol02}) and, (\ref{LTB}), then rewrite the metric in terms of $R$ as follows
\begin{equation}\label{Sol3}
	R\equiv e^{\phi/2}=\bigg[r^{\frac{3}{2}} (t_0-t)^{\frac{1-k}{2}}-\bigg(\frac{3}{2}\sqrt{2M}+\sqrt{6\pi\rho_0}  r^{\frac{3}{2}}\bigg)(t_0-t)^{\frac{1+k}{2}}\bigg]^{\frac{2}{3}} .
	\end{equation}
	
	In this regard, there are some special cases based on Eqs. (\ref{LTB}) and, (\ref{Sol3}) as follows:

\begin{itemize}
\item
$p_0\ne0$, $\rho_0\ne0$ and, $M\ne0$: black hole solution in the Quintom dominated universe;

\item
$p_0\ne0$, $\rho_0\ne0$ and, $M=0$: Quintom dominated cosmology;

\item
$p_0=0$, $\rho_0\ne0$ and, $M\ne0$: black hole solution in a dust dominated universe with $\rho_0=\rho_d a^3$, where, $\rho_d$ and $a$ are dust density and scale factor of the universe, respectively. Therefore, the metric of Eq. (\ref{Sol3}) turns into

\begin{equation}\label{Sol4}
	R=\bigg[r^{\frac{3}{2}}+\Big(\frac{3}{2}\sqrt{2M}+\sqrt{6\pi\rho_0}r^{3/2}\Big)t\bigg]^{\frac{2}{3}} ;
	\end{equation}

\item
$p_0=0$, $\rho_d=0$ and, $M\ne0$: Schwarzschild solution;

\item
$p_0=0$, $\rho_d\ne0$ and, $M=0$: dust dominated cosmology.

\end{itemize}
	
To investigate the apparent horizons of the cosmological LTB black hole immersed in a Quintom dominated universe we rewrite Eq. (\ref{LineElement}) based on the Schwarzschild notation

	\begin{equation}\label{Metric}
	ds^2=-(1-X^2) dt^2+dx^2+2Xdtdx+x^2 d\Omega^2 ,
	\end{equation}		
where
	\begin{equation}\label{partialx}
	x\equiv e^{\phi/2} \quad \textrm{and} \quad  X\equiv\frac{\partial{x}}{\partial{t}}.
	\end{equation}
 To find the apparent horizons, we benefit the new time coordinate like

	\begin{equation}
	dT=\bigg(dt+\frac{X}{1-X^2}dx\bigg) L^{-1} ,
	\end{equation}
where $L$ is a total differential that is a function of time and coordinate and therefore is not a constant. As a result, the metric of the cosmological LTB black hole in a Quintom dominated universe turns into

	\begin{equation}\label{mainmetric}
	ds^2=-(1-X^2) L^2 dT^2+\frac{1}{1-X^2} dx^2+x^2 d\Omega^2 .
	\end{equation}
	
\begin{figure}[ht]
		\centering
		\includegraphics [height=7 cm,width=9 cm] {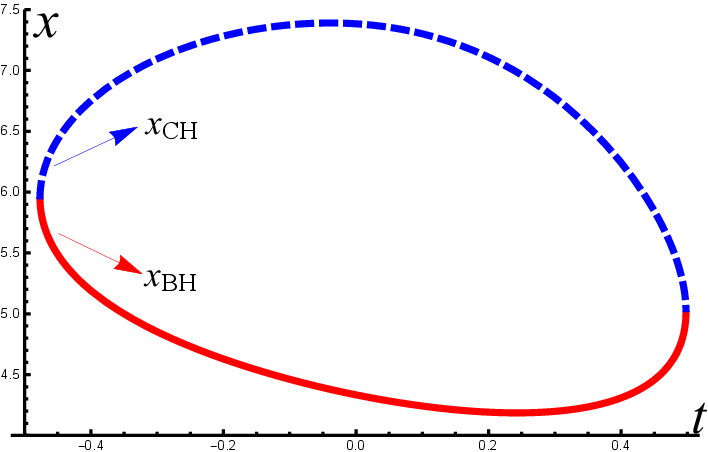}
		\caption{\scriptsize{The behavior of the cosmological and black hole apparent horizons versus time in blue curve (dashed line) and red curve (solid line), respectively. Plot has been depicted with fixed mass, $M=1$, while $\rho_0=0.0002$, $p_0=0.001$, and $t_0=1$.}}
		\label{figure_1}
		\end{figure}

	To calculate the apparent horizons, $x_{_{H}}$, we should find the roots of $\chi\equiv1-X^2=0$ which is equivalent to the following expression 	
	\begin{equation}\label{Hequation}
	1-\frac{4 \left(\sqrt{\frac{3 \pi \rho}{2}} r^{3/2} (1+k) (1-t)^{\frac{k-1}{2}}-\frac{1}{2} r^{3/2} (1-k) (1-t)^{-\frac{1+k}{2}}+\frac{3 (k+1) \sqrt{M} (1-t)^{\frac{k-1}{2}}}{2 \sqrt{2}}\right)^2}{9 \left(-\sqrt{6 \pi \rho} r^{3/2} (1-t)^{\frac{k+1}{2}}+r^{3/2} (1-t)^{\frac{1-k}{2}}-\frac{3 \sqrt{M} (1-t)^{\frac{k+1}{2}}}{\sqrt{2}}\right)^{2/3}}=0 ,
	\end{equation}
while we put $t_0=1$ in Eq. (\ref{partialx}). Therefore, substituting $r^{3/2}$ in terms of $x$, Eq. (\ref{Hequation}) would be an equation with six roots, some of which are the location of apparent horizons in this cosmological background. Setting $M=1$ and finding a numerical solution, we conclude that the second and third roots of the Eq. (\ref{Hequation}) are real and match with the boundary condition as we have illustrated them in Fig. \ref{figure_1}. The second root is the black hole apparent horizon, $x_{BH}$, and the third one is the cosmological apparent horizon, $x_{CH}$. There is a certain time in the past when the two horizons were coincided. Also, there is a certain time before the Big Rip when the two horizons will coincide again, and the naked singularity will be leftover. After creation of the horizons, with passing time, the size of the cosmological LTB black hole horizons in the Quintom universe evolves in such a way that the cosmological apparent horizon size (blue dashed curve) first increases and then decreases, while the black hole apparent horizon size (red solid curve) first decreases and then increases. It seems that the black hole horizon shrinking is due to the phantom component in this setup.

	\section{Thermodynamics of Cosmological LTB Black Hole in a Quintom Universe}
	Firstly, a brief description of how Hawking radiation works is explained in what follows. According to the quantum field theory, the vacuum is a complex entity of virtual particles that are continuously created, interacted, and then annihilated. In general, a vacuum is stable; but the presence of external fields makes it possible for the particles to become real. We suppose a static gravitational field with the Killing vector field $\xi^{\alpha}$. The particles' energy created in this field is equal to $\omega=-p_{\alpha}\xi^{\alpha}$, where $p^{\alpha}$ is four-momentum of the the particle and it is null for a massless particle. Whenever the virtual pair particle is created inside the horizon, the virtual particle with positive energy can tunnel throughout the horizon. Also, whenever the virtual pair particle is created outside the horizon, the virtual particle with negative energy can tunnel into the horizon. In both cases, the black hole absorbs the particle with negative energy, therefore, the mass of the black hole decreases; while the particle with positive energy escapes to infinity, and the observer detects it as Hawking radiation. Because the particle can classically fall into the black hole horizon, its action is real. For a particle that goes out the horizon of the black hole, the action becomes complex and the tunneling rate is determined by the imaginary part of the action. The transmission rate, $\Gamma$, which is equal to the probability of emission devided by the probability of absorption of particles, is related to the imaginary part of the action on one side and to the temperature on the other side, as follows
	\begin{equation}\label{Gamma}
	\Gamma=\frac{P_{em}}{P_{abs}}\sim\exp(-\beta\omega)\sim\exp(-2 \mathrm{Im} S)
	\end{equation}
where $\beta^{-1}$ is known as the temperature of the black hole. This explanation obliges us to calculate the imaginary part of the action to obtain the temperature of the black hole by quantum tunneling of the particles. There are two methods to calculate the imaginary part of the action: the Hamilton-Jacobi method \cite{99Sri} and the Parikh-Wilczek method \cite{00Par, 04Par}. The only noteworthy point remains that we are dealing with dynamic black holes instead of stationary ones.

In the cosmological context, a spherically symmetric black hole with a dynamical horizon cannot produce pure Hawking particle-antiparticle pairs, as this would break the principle of energy conservation and causes the apparent horizon to become spacelike \cite{15-Ellis}. In other words, the apparent horizon of any dynamical spacetime must lie inside the event horizon, and any virtual particle pairs created by the vacuum cannot escape and must fall back into the primordial black holes (PBHs). When we deal with fully dynamical metric, Hawking's quantum field theory approach to black hole radiation \cite{Davies-book} cannot be applied, as it is only suitable for late-time stationary black holes and cannot calculate the thermal aspect of Hawking radiation. Alternatively, new approaches \cite{99Sri, 00Par, 04Par} have been developed to calculate Hawking radiation in dynamical backgrounds. These approaches are based on the semiclassical approach using adiabatic vacuum in quantum field theory in curved spacetime, and suggest that radiation is likely emitted from the neighborhood of the apparent horizons rather than near the event horizon. In the case of dynamical black holes, universal definitions such as the black hole horizon and its surface gravity must be redefined based on local physics bases. The most important definitions are trapping horizon, which are introduced by Hayward \cite{94Hay}, and Kodama vector \cite{80Kod}. We are not going to explain these definitions here, but one can find some useful information on them in Refs. \cite{12Van, 15Far, 08Gou}.
Our strategy in what follows is to apply the Hamilton-Jacobi and Parikh-Wilczek methods separately to the cosmological LTB black hole in a Quintom dominated universe with the related definitions for the dynamical black holes.

\subsection{The Hamilton-Jacobi Method}
	
	The Hamilton-Jacobi equation for the cosmological LTB black hole in Quintom universe based on the metric (\ref{Metric}) is
	
	\begin{equation}\label{Action1}
	\chi (\partial_r S)^2-2X \omega (\partial_r S)-\omega^2=0 ,
	\end{equation}
where $S$ is the action and $\omega$ is the energy of a tunneling particle. We note that as before, $\chi$ is defined as $\chi\equiv1-X^2=0$ where $X\equiv\frac{\partial{x}}{\partial{t}}$ and $r$ is the comoving radial coordinate. The invariant particle energy is determined based on the Kodama vector, $K=(1,0,0,0)$, as follows

	\begin{equation}\label{Energy}
	\omega=-K^i \partial_i S=-\partial_t S .
	\end{equation}

It is important to note that Eq. (\ref{Action1}) contains both $r$ and $t$ since $\omega$ as the particle's energy is defined by the Kodama vector based on the time differentiation of the action.

Choosing the solution of the Eq. (\ref{Action1}) with positive radial momentum, we have

	\begin{equation}\label{Action2}
	\partial_r S=\frac{\omega X}{\chi} (1+O(\chi)) .
	\end{equation}
	
Therefore, $\partial_r S$ has a pole at the horizon. On the other hand, the action can be written as the sum of a real term and an imaginary term as follows

	\begin{equation}\label{Action3}
	S=\int{(dr \partial_r S+dt \partial_t S)}=\int{(dr \partial_r S+\frac{1}{2} \omega)} .
	\end{equation}
	
To calculate the imaginary part of the action which the first term contains it, we expand $\chi$ at the horizon as follows

 	\begin{equation}\label{chiexpand1}
	\chi\simeq\dot{\chi}\partial t+\chi' \partial x ,
	\end{equation}
where $\simeq$ means the approximation on the horizon and, $\partial x=x-x_H$. Also, from the metric (\ref{Metric}), outward null radial path crossing the horizon gives the result

	\begin{equation}\label{nullradial}
	\partial t=-(\frac{1}{2}X)\big\vert_{_{H}} \partial x .
	\end{equation}

Substituting Eq. (\ref{nullradial}) into Eq. (\ref{chiexpand1}), we conclude

	\begin{equation}\label{chiexpand2}
	\chi=\big(\chi'-\frac{1}{2X}\dot{\chi}\big)\bigg\vert_H (x-x_{_{H}})+... =2\kappa_{_{H}} (x-x_{_{H}})+O((x-x_{_{H}})^2) ,
	\end{equation}
where
	\begin{equation}\label{kappa}
	\kappa_{_{H}}=\frac{1}{2} \Box \: r\vert_{_{H}}=\frac{1}{2X^2} \big(\chi'-\frac{1}{2X}\dot{\chi}\big)\bigg\vert_H ,
	\end{equation}
is the dynamical surface gravity. Substituting Eq. (\ref{chiexpand2}) into Eq. (\ref{Action2}) and then in Eq. (\ref{Action3}), it is possible to calculate the imaginary part of the action using the Feynman's prescription as follows
	
	\begin{equation}\label{ImSHJ}
	\mathrm{Im} S=\mathrm{Im} \int {\partial_r S dr}=\mathrm{Im} \int{\frac{\omega X}{2 \kappa_{_{H}} (x-x_{_{H}}-i\epsilon)}dx}=\frac{\pi \omega_{_{H}}}{\kappa_{_{H}}} .
	\end{equation}

Finally, ussing Eq. (\ref{Gamma}) we can find the temperature of the cosmological LTB black hole immersed in Quintom universe as follows

	\begin{equation}\label{THJ}
	T=\beta^{-1}=\frac{\kappa_{_{H}}}{2\pi}.
	\end{equation}
	\\

\subsection{The Parikh-Wilczek Method}

	 Our approach is based on the quantum tunneling of the particles from the apparent horizon. We apply the null geodesics method which is well-known as the Parikh-Wilczek method~\cite{00Par}. Actually, the method describes the Hawking radiation by the pair of particle-antiparticle production near the horizon and the escape of the particle to infinity through the quantum tunneling process. The tunneling particle rate is related to both the imaginary part of the action and the temperature inverse. Therefore, calculations start with calculating the imaginary part of the action for a particle that is moving from an initial state at $x_{in}$ to the final state at $x_{out}$ as follows
	
	\begin{equation}\label{ImS}
	\mathrm{Im}S\equiv \mathrm{Im}\int
	E\:dt=\mathrm{Im}\int_{x_{in}}^{x_{out}}
	p_{x}\:dx=\mathrm{Im}\int_{x_{in}}^{x_{out}}\int_0^{p_{x}}\:d\tilde{p_x}\:dx,
	\end{equation}
where $x_{in}=x_{_{H}}-\epsilon$ and $x_{out}= x_{_{H}}+\epsilon$. Also, in what follows $\tilde{\omega}$ is the energy of the particle and we suppose this as a self interaction. With Hamilton equation, $dp_x=\frac{dH}{\dot{x}}$, Eq. (\ref{ImS}) changes to the following form

	\begin{equation}\label{ImS2}
	\mathrm{Im} S=\mathrm{Im}\int_{x_{in}}^{x_{out}}\int_{M}^{M-\tilde{\omega}}\frac{dH}{\dot{x}}\:dx=-
	\mathrm{Im} \int_0^{\tilde{\omega}}\int_{x_{in}}^{x_{out}}\frac{dx}{\dot{x}}\:d\omega.
	\end{equation}

We consider the lightlike geodesics for massless particles' tunneling regarded to the metric of Eq. (\ref{Metric}) (known as the Painlev\'{e}-Gullstrand like coordinate), we have

	\begin{equation}
	\dot{x}^2 +2\sqrt{1-\chi}\:\dot{x}-\chi=0.
	\end{equation}

As a result, we find the outgoing and ingoing trajectories as follows

	\begin{equation}\label{rdot}
	\dot{x}=\pm1-\sqrt{1-\chi}\,,
	\end{equation}
which gives $\dot{x}\simeq\frac{\chi}{2}$ for plus sign (outgoing trajectories). Substituting Eq. (\ref{rdot}) into Eq. (\ref{ImS2}), the imaginary part of the action for massless outgoing particles is given by

	\begin{equation}\label{PWImS}
	\mathrm{Im} S=-\mathrm{Im}\int_0^\omega\int_{x_{in}}^{x_{out}}\frac{2dx\:d\tilde{\omega}}{\chi}\,.
	\end{equation}

We put $\chi$ from Eq. (\ref{chiexpand2}) into Eq. (\ref{PWImS}), therefore, we can calculate the imaginary part of the action by Parikh-Wilczek method as follows

	\begin{equation}\label{PWImS2}
	\mathrm{Im} S=\int_0^\omega \frac{2 \pi \:d\tilde{\omega}}{2 \kappa_{_{H}}}=\frac{\pi \omega_{_{H}}}{\kappa_{_{H}}}\,.
	\end{equation}
	
As a result, the temperature with null geodesics approach will be the same which we obtained with Hamilton-Jacobi method in Eq. (\ref{THJ}). We expected the same outcome regardless of the calculation method since we expect the infinity observer to detect a certain temperature.

\subsection{Non-Thermal Spectrum}
	
	After the discovery of the thermal Hawking radiation, the information paradox has been discussed \cite{92Per, 93Pag}. Afterward, a criterion for calculating the correlation between radiation modes was proposed as follows~\cite{05Arz,Noz08}
	\begin{equation}\label{Corr}
	\zeta{(\omega_1+\omega_2;\omega_1,\omega_2)}=\ln{[\Gamma(\omega_1+\omega_2)]}-\ln{[\Gamma(\omega_1)	\Gamma(\omega_2])},
	\end{equation}
here $\zeta$ is the correlation function and $\omega_{1,2}$ are the tunneling particles' energy. Actually, Eq. (\ref{Corr}) lets us to know whether the probability of tunneling of two particles with energies $\omega_1$ and $\omega_2$ is the same as the probability of tunneling of one particle with energy $\omega_1+\omega_2$ or not. If the correlation between emitted modes is not zero, it means the radiation deviates from pure thermal radiation. Regarding Eq. (\ref{Gamma}), one can find that the transmission rate is related to the imaginary part of the action, and regarding Eq. (\ref{Corr}), the existence of a correlation between the emitted modes is obvious. Actually, we think that it is an important effect of the presence of the Quintom field in the environment of the black hole that causes this correlation between the emitted modes.

	\section{Evolution of Thermodynamic Features of Cosmological LTB Black Hole}
	
\begin{figure}[ht]
		\centering
		\includegraphics [height=7 cm,width=9 cm] {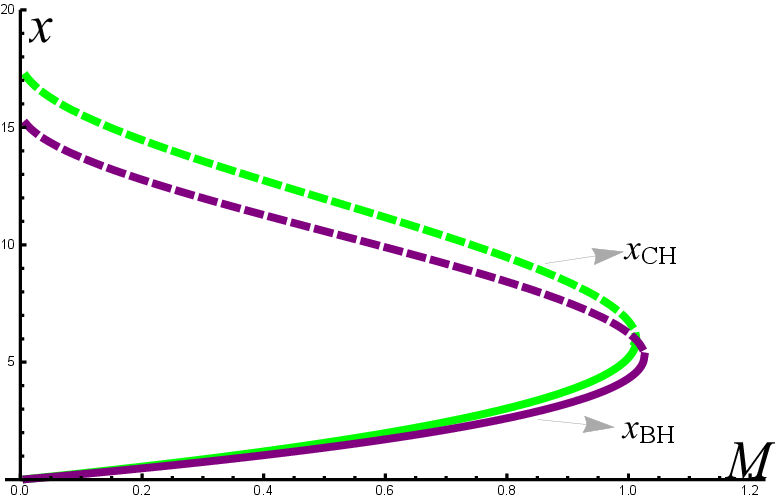}
		\caption{\scriptsize{The behavior of the cosmological and black hole apparent horizons versus the mass. Plot has been depicted with fixed time: $t=-0.4$ for the green curve and $t=+0.4$ for the purple curve. Solid lines show the black hole apparent horizons and dashed lines show the cosmological apparent horizons with $\rho_0=0.0002$, $p_0=0.001$, and $t_0=1$.}}
		\label{figure_2}
		\end{figure}

	We probed the time evolution of the horizons in the previous sections. In this section, we intend to investigate the effect of time evolution on the thermodynamics of the cosmological LTB black hole immersed in a Quintom universe. In other words, first of all, we obtain the apparent horizons in terms of the mass and derive the equation for temperature versus the mass of the black hole. Then, we evaluate the black hole temperature behavior in some cosmic epochs. This is important for us to answer the question whether the LTB black hole in a Quintom universe evaporates in the same way in all cosmic epochs or the time is an essential component that affects Hawking radiation and the black hole remnant. We have to find the apparent horizons from Eq. (\ref{Hequation}), but contrary to the previous section, here we want to fix the time and obtain an explicit expression in terms of the mass of the black hole. To describe precisely, if we consider a fixed time, there is a critical mass in which two apparent horizons coincide. As we illustrate in Fig. \ref{figure_2}, whatever the mass of the black hole is less than the critical mass, the two horizons are far away from each other; actually, the black hole horizon becomes smaller and the cosmological horizon becomes larger.
	
In order to obtain an explicit equation for the temperature in terms of the mass, first of all, we need the explicit expressions for the cosmological and black hole apparent horizons radii. These radii can be obtained via Eq. (\ref{Hequation}). The third root of the Eq. (\ref{Hequation}) is the cosmological apparent horizon, $x_{CH}$. Applying the self-gravitating shells \cite{95Kra}, we put $M-\omega$ instead of $M$ in $x_{CH}$. In this manner, we gain the cosmological apparent horizon after the particle tunneling, $x_{out}$ in Eq. (\ref{ImS2}). Selecting the outgoing trajectories from Eq. (\ref{rdot}), expanding $\dot{x}$ on the horizon, applying the residue calculus and expanding the result in terms of $\omega$, finally we obtain the imaginary part of the action as follows
	\begin{equation}\label{ImSfinal}
\mathrm{Im} S=\int_0^\omega \Big[\frac{320.1  x_{CH}^2}{ x_{CH}^3+22.9  x_{CH}^{3/2} \sqrt{M}-208.1 M}+O(\omega,\omega^2,...)\Big] d\omega.
	\end{equation}
The existence of the higher-order terms of $\omega$ proves the non-thermal nature of the radiation which we explained previously. Regarding Eq. (\ref{Gamma}), to calculate the temperature, we need to keep the coefficient of $\omega$ in the result of Eq. (\ref{ImSfinal}). As a result, we neglect higher-order terms of $\omega$ in this step and calculate the imaginary part of the action for a massless particles' tunneling. After that, based on Eq. (\ref{Gamma}), we find the temperature of the cosmological apparent horizon of the cosmological LTB black hole immersed in a Quintom universe as follows
\begin{equation}\label{TCH-0.4}
T_{CH}\bigg\vert_{t=-0.4}=\frac{1}{4\pi \beta}=\frac{0.000248569 \left(22.9 x_{CH}^{3/2} \sqrt{M}+x_{CH}^3-208.1 M\right)}{x_{CH}^2}.
	\end{equation}
In the same way, the temperature of the black hole apparent horizon of the cosmological LTB black hole immersed in a Quintom universe is as follows
\begin{equation}\label{TBH-0.4}
T_{BH}\bigg\vert_{t=-0.4}=\frac{1}{4\pi \beta}=\frac{0.000237356 \left(22.1 x_{BH}^{3/2} \sqrt{M}+x_{BH}^3-237.6 M\right)}{x_{BH}^2}.
	\end{equation}
	
\begin{figure}[ht]
		\centering
		\includegraphics[height=4 cm,width=5 cm]  {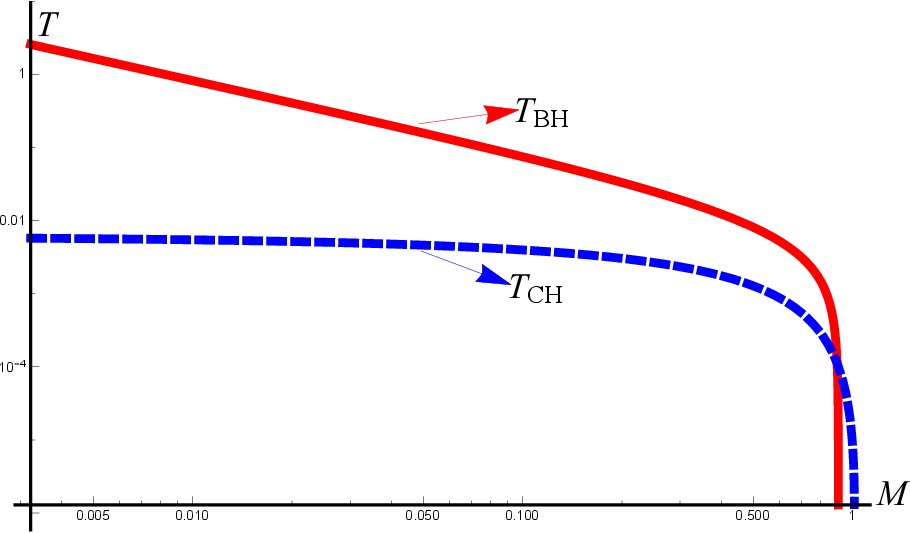}
		\hspace*{0.1 cm}
		\includegraphics[height=4 cm,width=5 cm] {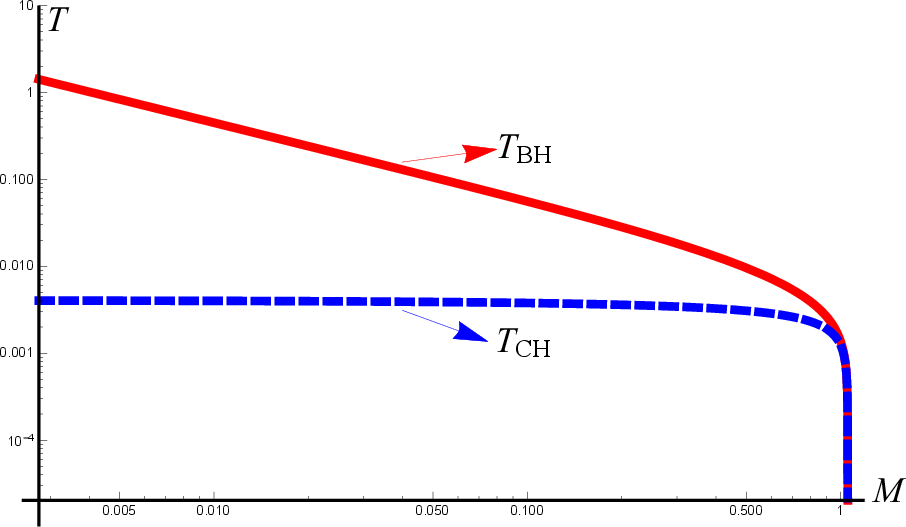}
		\hspace*{0.1 cm}
		\includegraphics[height=4 cm,width=5 cm] {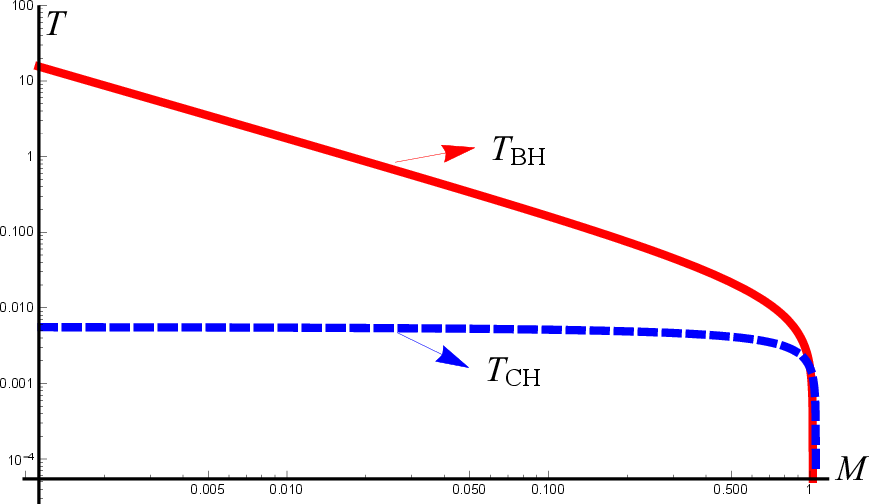}
		\caption{\scriptsize{The behavior of the black hole and cosmological apparent horizons' temperatures versus the mass in three cosmic epochs. We consider the fixed time equal to $t=-0.4,0,+0.4$ from left to right. The Hawking temperature of the black hole apparent horizon is more than the Hawking temperature of the cosmological apparent horizon in approximately all epochs. Whatever the cosmological LTB black hole in a Quintom universe evaporates in the early universe, its final temperature is expected to be lower.}}
		\label{figure_3}
	\end{figure}

We repeat the same calculations for the black hole horizon and also for these two horizons at other times.
Eventually, we find the temperature of the cosmological and black hole horizons of the cosmological LTB black hole in a Quintom universe as shown in Fig. \ref{figure_3}. In the critical mass, when two horizons created, the temperature starts to rise from zero. Approximately, in all of the cosmic history, the temperature of the black hole's apparent horizon is more than the temperature of the cosmological apparent horizon for the cosmological LTB black hole in a Quintom universe. Actually, the word \emph{approximately} is a keyword here, especially for the beginning of the Hawking radiation. The three panels of Fig. \ref{figure_3} are qualitative in essence since are drawn with some approximations and also all constants to be unity. The apparent horizon of black hole is always smaller than that of the universe; the main reason for the temperature of the black hole to be \emph{approximately} always higher than that of the universe. On the other hand, by comparing equations Eq. (\ref{TCH-0.4}) and Eq. (\ref{TBH-0.4}), we see that a smaller coefficient for the first term and a larger coefficient for the mass of the black hole with a minus sign may cause the temperature of the black hole horizon to be lower than the temperature of the cosmological horizon in some subspaces of the model parameter space, especially in the initial moments of the Hawking radiation. Conceptually, it may reflect the non-equilibrium situation in the first steps of the Hawking radiation emission. In another words, at the beginning steps of formation of the two horizons and Hawking radiation, the temperature of the cosmological horizon may be higher than the black hole temperature. But, after a short time, by the flow of energy between the two horizons via Hawking radiation, the two horizons attain the same temperature. Continuing to radiate via Hawking radiation, the temperature of the black hole horizon would be higher than the cosmological one as expected.

Also, there is a certain mass in which the two temperatures are the same. Comparing different epochs, at the time far from the Big Rip, it is predicted that the temperature of the cosmological LTB black hole immersed in a Quintom universe would be stopped at a lower temperature. In other words, in epochs closer to the Big Rip, for the cosmological LTB black hole in a Quintom universe, higher Hawking temperatures are expected in the final stage of the evaporation.\\
	
\begin{figure}[ht]
		\centering
		\includegraphics[height=5 cm,width=7 cm]  {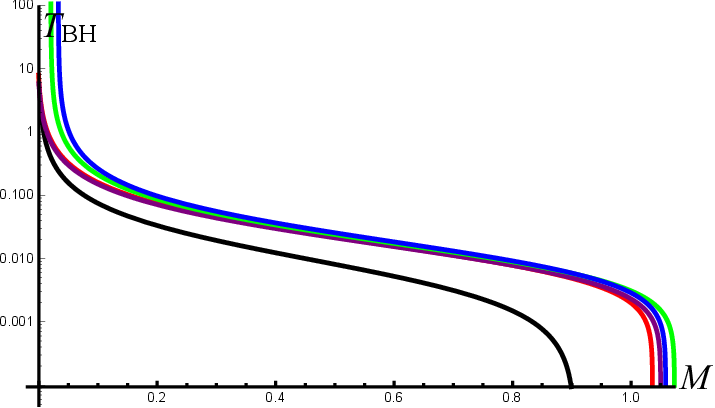}
		\hspace*{1cm}
		\includegraphics[height=5 cm,width=7 cm] {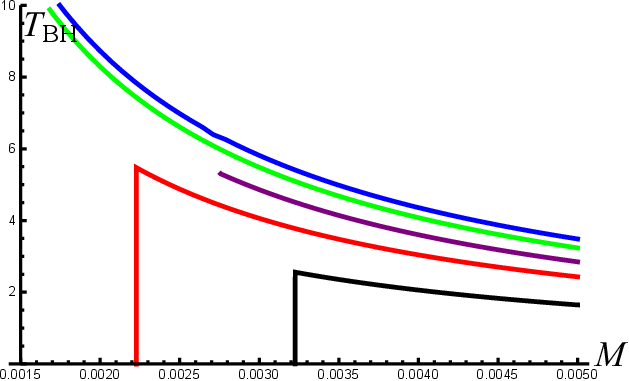}
		\caption{\scriptsize{The behavior of the black hole apparent horizon temperature versus the mass in some cosmic epochs. The left panel shows the universal behavior while the right panel shows the final stage of the evaporation in more details. We put fixed times $t=-0.4,-0.2,0,+0.2,+0.4$ from bottom to top. The temperature of the black hole horizon in the early universe falls into zero and the remnant with baryonic or dark energy content remains.}}
		\label{figure_4}
	\end{figure}

\begin{figure}[ht]
		\centering
		\includegraphics[height=5 cm,width=7 cm]  {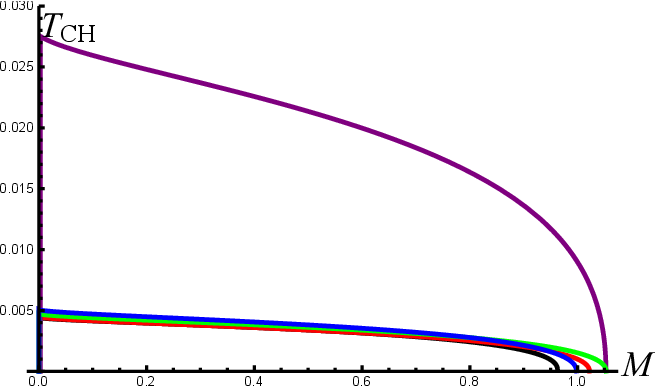}
		\hspace*{1cm}
		\includegraphics[height=5 cm,width=7 cm] {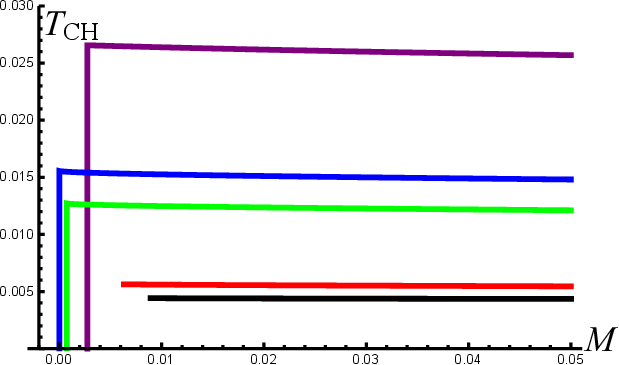}
		\caption{\scriptsize{The behavior of the cosmological apparent horizon temperature versus the mass in some cosmic epochs. The left panel shows the universal behavior while the right panel shows the final stage of the evaporation in more details. We put fixed times $t=-0.4,-0.2,0,+0.2,+0.4$ from bottom to top. The temperature of the cosmological horizon in the early universe is expected to reach a finite temperature and in the late time it is expected to fall into zero.}}
		\label{figure_5}
	\end{figure}

Moreover, we have illustrated Hawking temperature of the black hole apparent horizon and cosmological apparent horizon in some cosmic epochs in Figs. \ref{figure_4} and \ref{figure_5}, respectively. In these figures, the left panels represent the universal behavior of temperature and the right panels indicate the final stage of the evaporation in more detail. Actually, the results of the final stage of evaporation are interesting in some aspects; In cosmic epochs far from Big Rip, decreasing the mass, the cosmological horizon's temperature is expected to be constant while the black hole horizon's temperature first increases and then suddenly falls into zero. Conversely, in cosmic epochs close to the Big Rip, decreasing the mass, the cosmological horizon's temperature suddenly falls into zero and the black hole horizon's temperature is expected to increase slightly. The interesting point is the probability of the remnant formation. Indeed, we conclude if the cosmological LTB black hole in a Quintom universe evaporates in the early universe, the final remnant's content would be the baryonic matter. While, if it evaporates in the epochs close to the Big Rip,  the final remnant's content probably would be a dark energy content like Quintom matter.

About the sudden and sharp drop in the right panels of Figs. \ref{figure_4} and \ref{figure_5}, as we have mentioned previously, this is a trace of existing non-zero mass remnant with zero temperature. If the black hole evaporates in the early universe, evaporation continues until the temperature of the black hole horizon reaches zero, and the stable remnant remains. Maybe, these remnant can be a candidate for the primordial black hole and even cold dark matter. On the other hand, if the black hole evaporates in the late universe, Phantom domination causes the Big Crunch or Big Chill. Therefore, we can consider the zero temperature of the outer horizon of the black hole related to the Phantom dominance of the universe, growing the cosmological horizon size and Big Crunch/Big Chill. Existence of a non vanishing mass remnant has been observed in black hole evaporation in the contexts such as a noncommutative black hole, a quantum corrected black hole and especially for a black hole embedded in a scalar field. Therefore, observation of a sudden drop here is a trace of a non-zero mass remnant with vanishing temperature~\cite{Eslam2020}.

	Finally, we note that the calculation of temperatures in this setup should make sense in some adiabatic approximation, when the concept of temperature itself makes sense. Indeed, the correlation between $\omega_1$ and $\omega_2$ modes in Eq. (\ref{Corr}) can give a measure of the deviation from equilibrium. Indeed, if the evolution of the apparent horizons is fast, one does not expect a notion of equilibrium temperature to exist.
	
	\section{Summary and Conclusion}

In this work we have probed the cosmological LTB black hole immersed in a Quintom
universe. First, we have introduced the related metric and illustrated the time evolution of the
black hole and the cosmological horizons. We have shown that there is a certain time in
the past where the two horizons were coincided and, there is a certain time before the Big
Rip where the two horizons will coincide. In this respect, we have noticed that the black hole horizon
shrinking is due to the phantom component in this quintom model universe. Afterwards, we have applied two methods of
tunneling particles from the horizons. Precisely speaking, we have calculated the Kodama
vector and surface gravity based on the dynamical black hole definitions. Then, we
calculated the temperature of the cosmological LTB black hole in a Quintom universe. We
concluded that both Hamilton-Jacobi and Parikh-Wilczek methods have the same result for the
temperature of this black hole as we expected the infinity observer to detect a specified
temperature. Besides, we have shown the existence of a correlation between the
emitted modes and non-thermal nature of the spectrum which could be an address to the information
loss problem. Then we have investigated the temperature of the black hole and
cosmological horizons of the LTB black hole immersed in a Quintom universe at some
cosmic time slices. We have concluded that for both horizons in all cosmic time, there is
a critical mass in which two horizons are created, and the temperatures start to rise from
zero. Also, approximately in all the cosmic history, the temperature of the black hole's apparent horizon is
more than the temperature of the cosmological apparent horizon. On the other hand, in
epochs closer to Big Rip, for the cosmological LTB black hole in the Quintom universe,
higher Hawking temperatures are expected in the final stage of evaporation. Moreover,
we have illustrated the final stage of evaporation for both horizons at some cosmic time
epochs in more detail. The remarkable result is on the final remnant's content of the black
hole in the cosmic time close or far from the Big Rip. Actually, we have concluded that
the remnant of the LTB black hole would be a baryonic matter in the early universe and
would be a dark energy like Quintom matter in the epochs close to the Big Rip.\\


{\bf Acknowledgement:}
We would like to appreciate Valerio Faraoni for insightful comments on the original draft of this manuscript. Also, the authors appreciate
the respectful referee for carefully reading the manuscript and insightful comments which boosted the quality of the paper considerably.\\

{\bf Data Availability Statement:} This manuscript has no associated data or
the data will not be deposited. [Authors comment: We have no further
data related to this work to be deposited since it is definitely a theoretical
study. All possible data are included in the present paper.]\\

{\bf Conflict of Interest:}
There is no conflict of interest regarding this manuscript.

\end{document}